\documentclass[aps,prc,twocolumn,showpacs,superscriptaddress,nofootinbib,preprintnumbers]{revtex4-1}

\usepackage{longtable}
\usepackage{graphicx}
\usepackage{dcolumn}
\usepackage{bm}
\usepackage{amsmath,amssymb}

\usepackage[normalem]{ulem}  
\usepackage[dvipdfmx]{color} 

\renewcommand\sout{\bgroup \color{red} \ULdepth=-.5ex \ULset}

\newcommand{\physdim}[1]{\hspace{1ex} \mathrm{#1}}

\begin{document}
	
	\preprint{YITP-17-118}
	
	
	\title{Structure of hadron resonances with a nearby zero of the amplitude}
	
	
	\author{Yuki~Kamiya}
	\email[]{yuki.kamiya@yukawa.kyoto-u.ac.jp}
	\affiliation{Yukawa Institute for Theoretical Physics, Kyoto University, Kyoto 606-8502, Japan}
	\author{Tetsuo~Hyodo}
	\email[]{hyodo@yukawa.kyoto-u.ac.jp}
	\affiliation{Yukawa Institute for Theoretical Physics, Kyoto University, Kyoto 606-8502, Japan}
	
	
	\date{\today}
	
	\begin{abstract}
We discuss the relation between the analytic structure of the scattering amplitude and the origin of an eigenstate represented by a pole of the amplitude.
If the eigenstate is not dynamically generated by the interaction in the channel of interest,
the residue of the pole vanishes in the zero coupling limit.
Based on the topological nature of the phase of the scattering amplitude, we show that the pole  must encounter with the Castillejo-Dalitz-Dyson (CDD) zero in this limit. 
It is concluded that the dynamical component of the eigenstate is small if a CDD zero exists near the eigenstate pole. 
We show that the line shape of the resonance is distorted from the Breit-Wigner form as an observable consequence of the nearby CDD zero.
Finally, studying the positions of poles and CDD zeros of the $\bar{K}N$-$\pi\Sigma$ amplitude, we discuss the origin of the eigenstates in the $\Lambda(1405)$ region.
\end{abstract}
	
	
	
	
	\maketitle
	

\section{Introduction}
Given the recent findings of many candidates of exotic hadrons~\cite{Olive:2016xmw,Chen:2016qju,Hosaka:2016pey},
it is an urgent task to establish a method to clarify the internal structure of hadrons.
An important but difficult aspect of this task is brought by the
unstable nature of hadrons, which sometimes
spoils the straightforward interpretation of physical quantities to characterize the hadrons~\cite{Kamiya:2015aea,Kamiya:2016oao,Guo:2015daa}.
For instance, the energy of a resonance is uncertain, because of the finite decay width.
To define the energy of the resonance unambiguously, one can utilize the pole of the analytically continued scattering amplitude~\cite{Taylor:1972st},
which represents the eigenenergy of the Hamiltonian with the outgoing boundary condition,
as in the same way with the stable bound state.
It is therefore desirable to pin down the clue to characterize the internal structure of a resonance within the  scattering amplitude.

One step in this direction has been made by the pole counting method~\cite{Morgan:1990ct,Morgan:1992ge}.
In a coupled-channel scattering system, the position of poles in different Riemann sheets (shadow poles~\cite{Eden:1964zz}) is used as a test of the internal structure. 
It is argued that the resonance pole is accompanied by a nearby pole in the 
different Riemann sheet if the state is not dynamically generated.
Although the pole counting method is a powerful tool to qualitatively investigate the structure of resonances, it is required to determine the pole positions away from the most adjacent Riemann sheet.

Alternatively, several studies focus on the Castillegio-Dalitz-Dyson (CDD) zero~\cite{Castillejo:1955ed,Chew:1961} which is defined as the zero of the scattering amplitude.\footnote{The CDD zero is often referred to as ``CDD pole'', as it represents the pole of the inverse amplitude. In this paper, to avoid the confusion with the eigenstate pole, we call it CDD zero.}
The general form of the hadron scattering amplitude including the CDD contribution is discussed in Refs.~\cite{Oller:1998zr,Meissner:1999vr}.
The effect of the CDD zero on the effective range expansion in the near threshold region is discussed in detail for the case of a single channel in Ref.~\cite{Baru:2010ww} and for the case of multicontinuum channels in Ref.~\cite{Hanhart:2011jz}.
In a recent study~\cite{Guo:2016wpy}, it is shown that the CDD zero accompanied by nearby $\pi\Sigma_c$ thresholds performs the crucial role to reproduce the mass and width of $\Lambda_c(2595)$.
In Ref.~\cite{Kamiya:2016oao}, the CDD zero contribution near the threshold energy region in the estimation of the compositeness is discussed.
While the importance of the CDD zero contribution is recognized, the direct relation between the structure of the state and the CDD zero is still unclear.

In this paper, we show that the distance between the eigenstate pole and the CDD zero is related to the structure of the state.
To reveal the origin of the eigenstate, we consider the zero coupling limit (ZCL), which is the limit of turning off the couplings among different coupled channels. 
By analyzing the behavior of CDD zeros and poles in the ZCL, we discuss the relation between the internal structure of the eigenstate and the existence of a nearby CDD zero.

\section{Zero coupling limit and origin of eigenstate}
Our aim is to clarify the dynamical origin of the eigenstate expressed by a pole of the coupled-channel scattering amplitude in a given partial wave.
For this purpose, we focus on one of the coupled channels, say channel $i$,
and consider the diagonal component of the scattering amplitude $\mathcal{F}_{ii}(E)$.
We would like to utilize the analytic structure of the scattering amplitude, namely,
the position of the CDD zero $\mathcal{F}_{ii}(E_{\mathrm{CDD}})=0$ relative to the eigenstate pole $\mathcal{F}_{ii}(E_{\rm pole})=\infty$.
In this study, we concentrate on $E_{\mathrm{CDD}}$ and $E_{\mathrm{pole}}$ in the most adjacent Riemann sheet to the physical real energy.
For later convenience, we recall the fact that $E_{\mathrm{pole}}$ is common to all the channel components, whereas the position of the CDD zero can be channel dependent.

To elucidate the origin of the eigenstate, we consider the zero coupling limit (ZCL)~\cite{Eden:1964zz,Pearce:1988rk,Cieply:2016jby},
where the off-diagonal couplings between different channels are switched off,
but the diagonal interactions are kept fixed. 
We can draw the trajectory
of the pole toward the ZCL, by gradually switching off the channel couplings.
While the pole exists in the all components of the coupled channel amplitude  with nonzero
channel couplings, in the exact ZCL, the pole remains only in the one of the components
and decouples from the others.\footnote{If a pole remains in two or more components in the exact ZCL, the degenerate eigenstates must exist.
In this case, there must be a symmetry which relates different channels. 
If the eigenstate is generated not by the interaction of a specific channel
but purely generated by the channel coupling effect, the pole cannot remain in any of the components.
In this case, the pole should move away to infinity as $|E_{\rm pole}|\to \infty$ in the ZCL.
In the following, we do not consider these special cases, and concentrate on the case where the eigenstate originates in the dynamics of one specific channel.}
It is natural to attribute the origin of the eigenstate to the dynamics of the channel
where the pole remains in the ZCL.

Thus, we can classify the behavior of the pole in a specific channel into two cases:
\begin{enumerate}
	\item The pole remains in the amplitude in the ZCL.
	\item The pole decouples from the amplitude in the ZCL.
\end{enumerate}
In the former case, the origin of the eigenstate is attributed to the dynamics of this channel.
The latter case is achieved by the vanishing of the residue of the pole and the eigenstate originates in the dynamics of the other channels.
In the following, we show that the CDD zero is closely related to this latter case.
Before addressing the general case, it is instructive to study two examples
in which the position of the eigenstate pole and the CDD zero can be calculated explicitly.

First, we consider a single-channel scattering problem coupled to a bare state, utilizing the nonrelativistic effective field theory introduced in  Ref.~\cite{Kamiya:2016oao}.
The effective field theory includes the fields $\psi$, $\phi$ and $B_0$, with the Hamiltonian given as
\begin{align}
H &=\int d^3\bm{x} \biggl[ \frac{1}{2 M} \mathbf{\nabla} \psi^\dagger \cdot\mathbf{\nabla} \psi +\frac{1}{2 m} \mathbf{\nabla} \phi^\dagger \cdot\mathbf{\nabla} \phi\notag\\
&\quad+ \frac{1}{2M_{0}} \mathbf{\nabla}  B_0^\dagger \cdot{\mathbf \nabla} B_0+ 
\omega_0 B_0^\dagger B_0,\notag\\
 &\quad+ g_0\left( B_0^\dagger \psi\phi + \phi^\dagger\psi^\dagger B_0 \right) +  v_{0} \psi^\dagger\phi^\dagger \phi\psi\biggr],\label{eq:H_int_single}
\end{align}
with $\hbar=1$. 
The fields $\psi$ and $\phi$ compose the scattering channel and the field $B_0$ expresses the discrete channel with a discrete energy level, e. g. compact quark state.
The exact on shell $T$-matrix in this system is obtained as
\begin{align}
t(E)  = \frac{v_0(E-\omega_0) +g^2_0}{(E-\omega_0)[1-v_0G(E)] -g_0^2 G(E)},\label{eq:T_single}
\end{align}
where $G(E)$ denotes the loop function of the scattering channel
\begin{gather}
G(E)\equiv \int\frac{d^3\bm{p}}{(2\pi)^3}\frac{1}{E-p^2/(2\mu)+i0^+},
\end{gather}
 and $\mu=Mm/(M+m)$.
The scattering  amplitude $\mathcal{F}(E)$ relates to the $T$-matrix as $\mathcal{F}(E) = - \mu t(E)/(2\pi)$.

We consider the case where the system has a discrete eigenstate. 
The zero of the denominator of Eq.~\eqref{eq:T_single} tells us the pole position of the amplitude $E=E_{\mathrm{pole}}$ is
\begin{align}
(E_{\mathrm{pole}}-\omega_0)[1-v_0G(E_{\mathrm{pole}})] -g_0^2 G(E_{\mathrm{pole}})=0. \label{eq:pole_condition_single}
\end{align}
The residue of the pole is calculated by $\lim_{E\rightarrow E_{\mathrm{pole}}}(E-E_{\mathrm{pole}})t(E)$ as 
\begin{align}
  \frac{v_0(E_{\mathrm{pole}}-\omega_0) +g_0^2}{1-v_0\left[G(E_{\mathrm{pole}})+(E_{\mathrm{pole}}-\omega_0)G^\prime(E_{\mathrm{pole}})\right]+g_0^2G^\prime(E_{\mathrm{pole}})}, \label{eq:res-single}
\end{align}
where $G^\prime(E)$ denotes the energy derivative of the function $G(E)$.

Let us examine the behavior of the pole $E_{\mathrm{pole}}$ in the ZCL. 
Because there is a bare state $B_0$ in addition to the scattering channel, the ZCL corresponds to the limit of  $g_0\rightarrow 0$.
In this case, Eq.~\eqref{eq:pole_condition_single} indicates that $E_{\mathrm{pole}}\rightarrow \omega_0$ or $E_{\mathrm{pole}}\rightarrow E_{\mathrm{0}}$ with $1-v_0G(E_{\mathrm{0}})=0$.
If the origin of the eigenstate is the interaction of the scattering channel,
the pole must remain in the amplitude at $E_{\mathrm{pole}}=E_0$ because the position of $E_{\mathrm{0}}$ is determined within the scattering channel. 
On the other hand, if the bare state is the origin of the eigenstate,
the pole behaves as $E_{\mathrm{pole}}\rightarrow \omega_0$. 
In the absence of degeneracy of the eigenstate,
we have  $1-v_0G(\omega_0)\neq0$, and therefore, the residue of the pole in Eq.~\eqref{eq:res-single} vanishes in the ZCL.

Next let us consider the CDD zero $E_{\mathrm{CDD}}$, defined as $\mathcal{F}(E_{\mathrm{CDD}})=0$. In the present case, $E_{\mathrm{CDD}}$  is determined by the zero of the numerator of the $T$-matrix in
Eq.~\eqref{eq:T_single}, so
the position of the CDD zero can be written as
\begin{align}
E_{\mathrm{CDD}} = \omega_0 -g_0^2/v_0.
\end{align}
In the ZCL ($g_0\rightarrow0$), we find $E_{\mathrm{CDD}}\rightarrow\omega_0$.
Thus, if the bare state is the origin of the eigenstate, both 
the pole and the CDD zero move toward $E=\omega_0$ by reducing the coupling $g_0$.
In the exact ZCL ($g_0 =0$), the pole and the CDD zero no longer exist because 
the zero of the numerator and denominator in Eq.~\eqref{eq:T_single} cancel out each other.
When the coupling $g_0$ is small,
the deviation $E_{\mathrm{CDD}} - \omega_0$ and $E_{\mathrm{pole}}- \omega_0$ are of the order of $g_0^2$. This means that the distance between the pole and the CDD zero $E_{\mathrm{pole}} - E_{\mathrm{CDD}}$ is also $\mathcal{O}(g_0^2)$.
Thus, we conclude that the pole and the CDD zero lie close to each other when the eigenstate originates in the bare state channel.

Next, we consider the system with two scattering channels.
We introduce fields $\psi_i$ and $\phi_i$ which consist of the scattering channel $i=1,2$.
The Hamiltonian is given as 
\begin{align}
H &=\int d^3\bm{x}\biggl[\sum_{i=1,2}\bigl[\frac{1}{2 M_i} \mathbf{\nabla} \psi_i^\dagger \cdot\mathbf{\nabla} \psi_i  
+\frac{1}{2 m_i} \mathbf{\nabla} \phi_i^\dagger  \cdot\mathbf{\nabla} \phi_i  \bigr]\notag \\
&\quad-\omega_\psi \psi_2^\dagger \psi_2 -\omega_\phi \phi_2^\dagger \phi_2+ \sum_{i,j=1,2} v_{0,ij} \psi_j^\dagger \phi_j^\dagger  \phi_i \psi_i\biggr].
\end{align}
Similarly to the previous problem, 
the on-shell $T$-matrix of channel 1 is obtained as 
\begin{align}
t_{11}(E) &= \frac{v_{0,11}[1-v_{0,22}G_2] +v_{0,12}^2G_2}{[1-v_{0,11}G_1][1-v_{0,22}G_2]-v_{0,12}^2G_1G_2}\label{eq:T-mat-coupled},
\end{align}
with 
\begin{align}
G_i(E) \equiv \int \frac{d^3\bm{p}}{(2\pi)^3}    \frac{1}{E- p^2/(2\mu_i) + \delta_{i,2}(\omega_\psi+\omega_\phi)+i0^+},\label{eq:loop_fcn_coupled}
\end{align}
with $\mu_i=M_i m_i /(M_i+m_i)$.
The scattering amplitude of channel $i$ is given as $\mathcal{F}_{ii}(E) = - \mu_i t_{ii}(E)/(2\pi)$.

Again, we consider the case with one eigenstate at $E=E_{\mathrm{pole}}$.
With Eq.~\eqref{eq:T-mat-coupled}, the conditions for $E_{\mathrm{pole}}$ is given as
\begin{align}
[(1-v_{0,11}G_1)(1-v_{0,22}G_2)-v_{0,12}^2G_1G_2]|_{E=E_{\mathrm{pole}}}=0.\label{eq:pole_condition_coupled}
\end{align}
In this model, the ZCL corresponds to the limit of $v_{0,12}\rightarrow 0$ where the pole behaves as 
$E_{\mathrm{pole}}\rightarrow E_{0,1}$ or $E_{0,2}$ with $1-v_{0,11}G_1(E_{0,1})=0$ and $1-v_{0,22}G_2(E_{0,2})=0$.
If the dynamical contribution of channel 1 is the origin of the eigenstate, 
the pole remains in the amplitude at $E_{\mathrm{pole}}=E_{0,1}$ in the ZCL.
If not, the pole moves toward $E_{0,2}$ which plays a similar role with the bare state energy $\omega_0$ in the previous example. 
It can also be shown that the residue of the pole vanishes at $E = E_{0,2}$ in this case.

The position of the CDD zero is determined from Eq.~\eqref{eq:T-mat-coupled} as
\begin{align}
v_{0,11}[1-v_{0,22}G_2(E_{\mathrm{CDD}})] +v_{0,12}^2G_2(E_{\mathrm{CDD}})=0. \label{eq:CDD_condition_coupled}
\end{align}
In the ZCL, 
namely, both $E_{\mathrm{CDD}}$ and $E_{\mathrm{pole}}$ move to $E_{0,2}$ if the dynamical contribution of channel 2 is the origin of the eigenstate.
As with the previous example, the distance $E_{\mathrm{pole}}-E_{\mathrm{CDD}}$ is of the order of the square of coupling constant $\mathcal{O}(v_{0,12}^2)$.

It is also instructive to comment on the realization of the pole counting rule~\cite{Morgan:1990ct,Morgan:1992ge} in this model.
In the pole counting method, one studies the position of the shadow pole which lies in the other Riemann sheet than that of the eigenstate pole~\cite{Eden:1964zz}.
It is conjectured in Refs.~\cite{Morgan:1990ct,Morgan:1992ge} that a shadow pole does not appear near the eigenstate pole of a dynamically generated state, and the existence of a nearby shadow pole indicates the hidden-channel origin of the eigenstate. 
In the present model, the Riemann sheet can be chosen by changing the sign of $i0^{+}$ in Eq.~\eqref{eq:loop_fcn_coupled}. 
The equation to determine the shadow pole is then obtained by modifying $G_{1}$ in Eq.~\eqref{eq:pole_condition_coupled}. 
Thus, if the eigenstate is generated in channel 2, near the ZCL, the shadow pole approaches $E_{0,2}$ and its deviation from $E_{0,2}$ is $\mathcal{O}(v_{0,12}^2)$. 
This means that the shadow pole appears near the eigenstate pole. Note that the shadow pole also encounters with a CDD zero in its own Riemann sheet, because the equation for the CDD zero~\eqref{eq:CDD_condition_coupled} does not depend on $G_{1}$. 
In this way, the shadow pole in this model behaves in accordance with the pole counting method.

Let us shortly summarize the discussion of this part.
We have studied the behavior of the pole and zero of the amplitude in the ZCL in two models.
In both cases,
we find that the pole representing an eigenstate and the CDD zero annihilate each other in the ZCL if the origin of the eigenstate is attributed to the dynamics of the hidden channel. 
It is also shown that the distance between the pole and the CDD zero is small
when the channel coupling is small.

\section{Origin of eigenstate and nearby CDD zero}
Here we show that the annihilation of the pole and zero is, in fact, a consequence of the general property of the scattering amplitude. Let $\mathcal{F}(z)$ be an analytically continued partial-wave scattering amplitude with the complex energy variable $z$. Because $\mathcal{F}(z)$ is a meromorphic function of $z$, for a counterclockwise closed contour $C$ on which $\mathcal{F}(z)$ is analytic, the argument principle leads to
\begin{align}
\frac{1}{2\pi } \oint_C dz \frac{d\arg \mathcal{F}(z)}{dz} 
=n_{Z}-n_{P} \equiv n_{C},
\label{eq:nC}
\end{align}
where the integers $n_{Z}$ and $n_{P}$ represent the number of zeros and poles of $\mathcal{F}(z)$ enclosed by $C$.\footnote{Here we assume that all the poles and zeros are simple. Singularities with multiplicity can appear only with a fine tuning of the parameters, and its existence is not stable against the small perturbation of the parameters.} For instance, if $C$ encloses a single pole, $n_{C}=-1$. If there is neither a pole nor a zero in $C$, $n_{C}=0$. Equation~\eqref{eq:nC} can be shown by rewriting the integrand as $(d\mathcal{F}(z)/dz)/i\mathcal{F}(z)$. The expression in Eq.~\eqref{eq:nC} indicates that $n_{C}$ is an integer because it is the topological invariant of $\pi_{1}({\rm U}(1))\cong \mathbb{Z}$. 

If we continuously vary the parameters of the system (e.g. reducing the channel couplings toward the ZCL), the poles and zeros of the scattering amplitude move continuously in the complex $z$ plane. As long as the contour $C$ is chosen not to intersect with the trajectories of poles and zeros, the value of $n_{C}$ cannot change along with the continuous variation of the parameters, due to its topological nature. This means that an abrupt transition from the amplitude with one pole in $C$ ($n_{C}=-1$) to nothing ($n_{C}=0$) is forbidden. In order for a pole to decouple, there must exist a zero in $C$ so that $n_{C}=0$ in total, and the pole and the zero must encounter to cancel out with each other, as we have observed in the examples. 

This feature of the scattering amplitude is particularly important for the fate of the pole in the ZCL.\footnote{The above argument can be applied to a single component of coupled-channel amplitude $F_{ii}(z)$ by choosing the integration contour $C$ to avoid the branch cuts.} As we have discussed, the pole decouples from the amplitude in the ZCL if the state is not dynamically generated in the channel of interest. Here we find that such pole must be accompanied by a nearby CDD zero. In other words, there is a simple rule to elucidate the origin of the eigenstate:
\begin{enumerate}
	\item If there is no CDD zero near a pole in $\mathcal{F}_{ii}(z)$, the eigenstate is dynamically generated in channel $i$.
	
	\item If a pole is accompanied by a nearby CDD zero, the origin of the eigenstate is not in channel $i$.
\end{enumerate}
With the scattering amplitude fitted to reproduce the experimental data, this rule can be directly utilized in order to determine the origin of the eigenstate by searching the positions of the poles and zeros of the amplitude.\footnote{Here we focus on the  zeros of the diagonal component of the scattering amplitude.
However, when the effect of the multichannel (multicomponent) is important, the production amplitude obtained in an experiment is the mixture of the different terms.
The position of the zero of this production amplitude can be different from that of the diagonal scattering amplitude.
Therefor, in order to discuss the structure of the eigenstates, it is necessary to extract the coupled channel scattering amplitude by detailed analysis of the experimental data and to search the position of zero of the diagonal scattering amplitude.}
We note that the position of the CDD zero can, in principle, be uniquely determined from the amplitude because of the uniqueness of the analytic continuation. Moreover, as we show below, the existence of the CDD zero has a physical consequence in the two-body spectrum. In this sense, the above rule can be used as a practical method to pin down the origin of an eigenstate from the observable quantity. 

The existence of the CDD zero near the pole causes yet another consequence for the near-threshold states.
The effective range expansion (ERE) is often introduced in the analysis of near-threshold energy region of the scattering amplitude.
However, the CDD zero near the threshold harms the convergence of the ERE as pointed out in Refs.~\cite{Baru:2010ww,Guo:2016wpy,Kamiya:2016oao}.
When the above case 2 is realized, because of the nearby CDD zero, the analysis with ERE may fail to describe the eigenstate pole.
Conversely, the failure of the ERE is an indication of the external origin of the eigenstate in accordance with the result in Ref.~\cite{Guo:2016wpy}.

Because the analytic structure of the scattering amplitude is utilized to determine the origin of the eigenstate,
the method constructed above has similarity with the pole counting method~\cite{Morgan:1990ct,Morgan:1992ge}.
However, there is one difference; while the CDD zero lies in the most adjacent Riemann sheet to the real axis as the eigenstate pole does, the shadow poles exist in Riemann sheets which are  not directly connected to the real axis. 
Thus, it is an advantage of the present method to utilize the position of the CDD zero which can be determined less ambiguously than those of the shadow poles.  
Moreover, the CDD zero can appear on the real axis, as in the example
in the next section, and also in the  $\pi\Sigma(I=0)$ scattering~\cite{Kamiya:2016oao}.
In terms of the phase shift $\delta$, the CDD zero corresponds to $\delta = \pi$,
which is indeed observed in $\pi\pi(I=0)$ scattering near the $\bar{K}K$
threshold~\cite{Pelaez:2015qba}.
In this way, for the CDD zero on the real axis, the analytic continuation of the
amplitude is no longer necessary, and the phase shift data directly
determines its location.

While our method draws qualitative conclusion on the origin of the eigenstate, one may want to discuss the structure quantitatively.
For this purpose, we can calculate Weinberg's $Z$ or the amount of the scattering state in the eigenstate wave function, the so-called compositeness $X$~\cite{Weinberg:1965zz,Baru:2003qq,Baru:2010ww,Hyodo:2011qc,Aceti:2012dd,Hyodo:2013nka,Hyodo:2013iga,Hyodo:2014bda,Sekihara:2014kya,Aceti:2014ala,Garcia-Recio:2015jsa,Guo:2015daa,Guo:2016wpy,Kamiya:2015aea,Sekihara:2015gvw,Kamiya:2016oao,Tsuchida:2017gpb}.
The quantity $Z$ is introduced as the field renormalization constant in Ref.~\cite{Weinberg:1965zz},
and is later extended for the study of the hadron resonances in Ref.~\cite{Baru:2003qq} by using the spectral density approach~\cite{Bogdanova:1991zz}.
In the recent study, it is found that the compositeness $X$ can be calculated by  the residue of the pole in the scattering amplitude in Ref.~\cite{Hyodo:2011qc,Aceti:2012dd,Guo:2016wpy}.
However, $Z$ or $X$ of the unstable state with a finite decay width becomes a complex number that is difficult to interpret physically, and it generally suffers from model dependence except for near-threshold states in $s$ wave~\cite{Kamiya:2015aea,Kamiya:2016oao}.
On the other hand, as in the pole counting rule, our method can only draw qualitative conclusions.
However, using this method, it is possible to discuss the state other than the $s$ wave and the state away from the threshold in a model-independent manner.
Therefore, since the two methods have different advantages, practically it is better to use both of them complementarily.

\section{Observable consequence of CDD zero near resonance}
In this section, we consider the influence of the existence of a CDD zero near the pole.
Again, we take an example of coupled-channel scattering model with the Lagrangian
in Eq.~(11) for illustration.
We consider the coupled-channel $N\Xi$ (channel 1) and
$\Lambda \Lambda$ (channel 2) scattering, in which a quasibound state ($H$-dibaryon)
can be generated by the attractive $N\Xi$ interaction as discussed in Ref.~\cite{Yamaguchi:2016kxa}.

Taking the model parameters as $v_{0,11} = -8.93\times 10^{-5}\mathrm{MeV}^{-2}$, $v_{0,12}=1.20\times 10^{-5}\mathrm{MeV}^{-2}$, $v_{0,22}=-3.53\times 10^{-5}\mathrm{MeV}^{-2}$, and  $\Lambda=300\physdim{MeV}$, we find an eigenstate at $E_{\mathrm{pole}}=-5-1i\physdim{MeV}$ measured from the threshold energy of the $N\Xi$ channel.
We find a CDD zero at $E_{\mathrm{CDD}}=-4\physdim{MeV}$ in the $\Lambda\Lambda$ amplitude.
This zero appears close to the physical pole, implying that the $\Lambda\Lambda$ dynamical component is not dominant in accordance with the $N\Xi$ molecule picture.

\begin{figure}[t]
\centering
  \includegraphics[width=75mm]{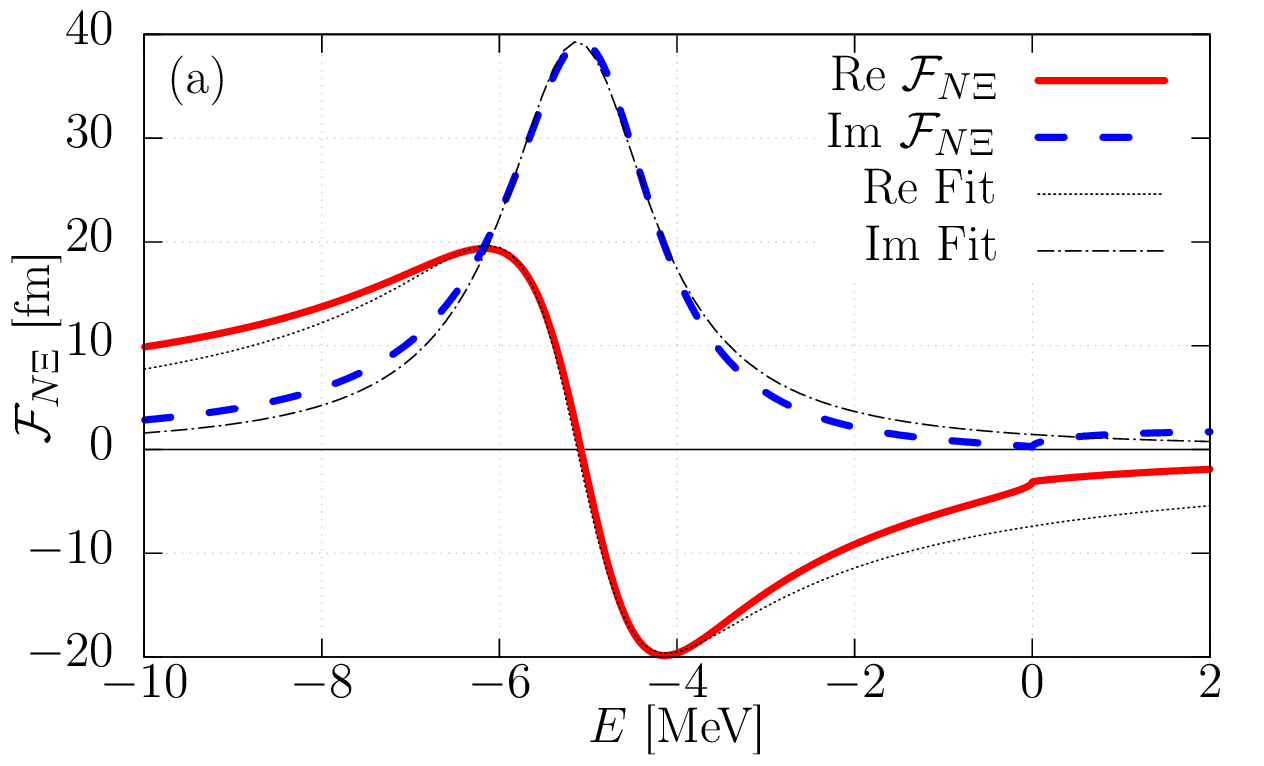}
  \includegraphics[width=75mm]{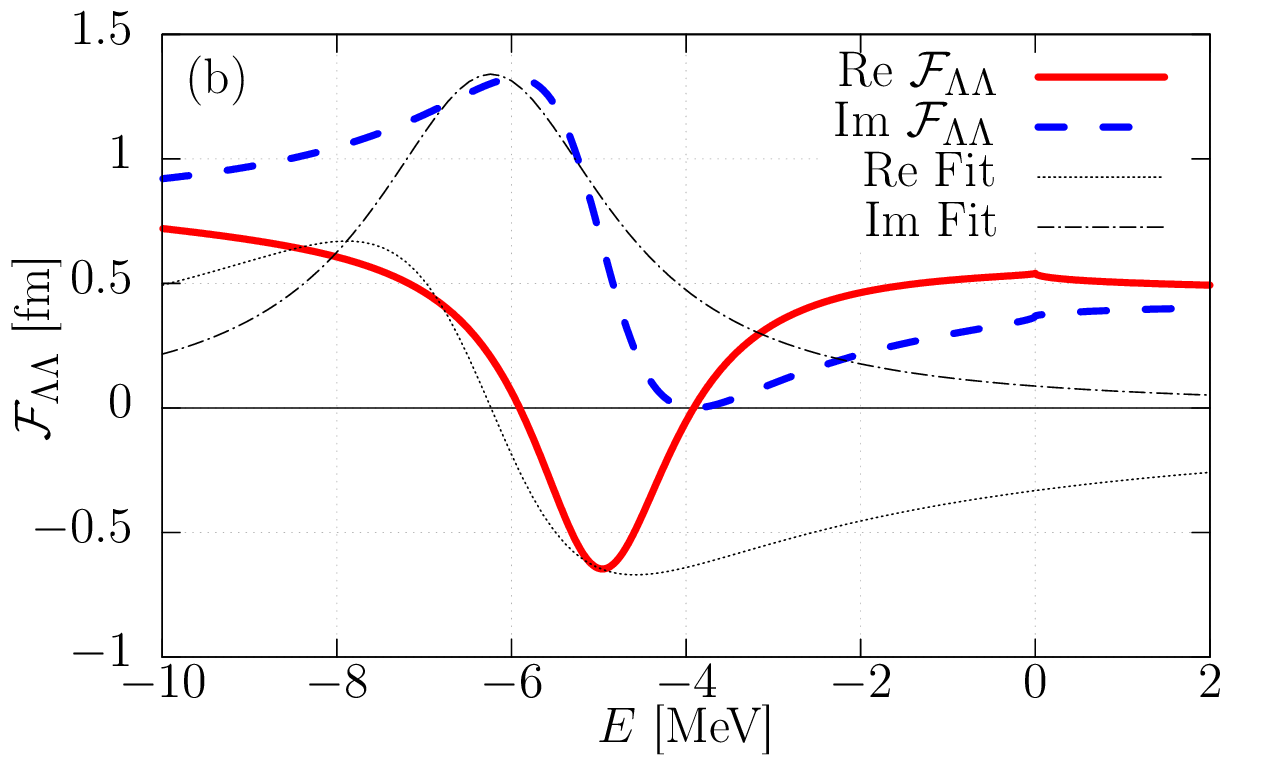}
   \caption{The $N\Xi\rightarrow N\Xi$ (a) and $\Lambda\Lambda\rightarrow\Lambda\Lambda$  (b) scattering amplitude near the $N\Xi$ threshold energy region. The solid line and dashed line denote the real part and imaginary part of the scattering amplitude, respectively. The dotted line and the dash-dotted line denote the real part and the imaginary part of the fitted amplitude by the Breit-Wigner function.}
   \label{fig:amp_H}
\end{figure}
The scattering amplitudes of the $\Lambda\Lambda$ and $N\Xi$ channel are plotted in Fig.~\ref{fig:amp_H}.
In both the amplitudes, a peak structure is observed below the $N\Xi$ threshold.
To see the distortion of the line shape, we fit these amplitudes around the peak position by the Breit-Wigner function $\mathcal{F}(E) =g^2 /(E-M_r+i\Gamma_r/2)$ as shown in Fig.~\ref{fig:amp_H}.
In the $N\Xi$ amplitude, the peak structure is well approximated by the Breit-Wigner function.
However, the line shape in the $\Lambda\Lambda$ amplitude is highly asymmetric and distorted from the Breit-Wigner form.
Such distortion is caused by the nearby CDD zero,
because both the real and imaginary part of the amplitude must vanish at $E=E_{\mathrm{CDD}}$.
This requirement shifts the peak position from the real part of the pole to the other side of the CDD zero and distorts the peak structure.
In general, the CDD zero can lie in the complex energy plane, but the distortion of the line shape can also occur in this case. 
This is because the zero lies in the most adjacent Riemann sheet to the physical real energy, which directly affects the line shape.

The limited applicability of the Breit-Wigner fit for the near-threshold states is pointed out in Ref.~\cite{Guo:2016bjq}.
In addition to this admonition,  the present study 
shows that we have to pay attention to the CDD zero around the resonance.
Even away from the threshold, the CDD zero may hamper the application of the Breit-Wigner fit. 

\section{Application to $\Lambda(1405)$}
Let us now consider the physical $I=0$ $\bar{K}N$ scattering to study the origin of the $\Lambda(1405)$ baryon resonance.
In the recent analyses of experimental data with next-to-leading order  chiral SU(3) dynamics~\cite{Olive:2016xmw,Ikeda:2011pi,Ikeda:2012au,Mai:2012dt,Guo:2012vv,Mai:2014xna}, 
it is shown that there are two poles in the $\Lambda(1405)$ region~\cite{Jido:2003cb}.
We denote the pole near the $\bar{K}N$ ($\pi\Sigma$) threshold energy as high-mass (low-mass) pole according to Ref.~\cite{Olive:2016xmw}.

\begin{figure}[t]
	\centering
	\includegraphics[width=75mm]{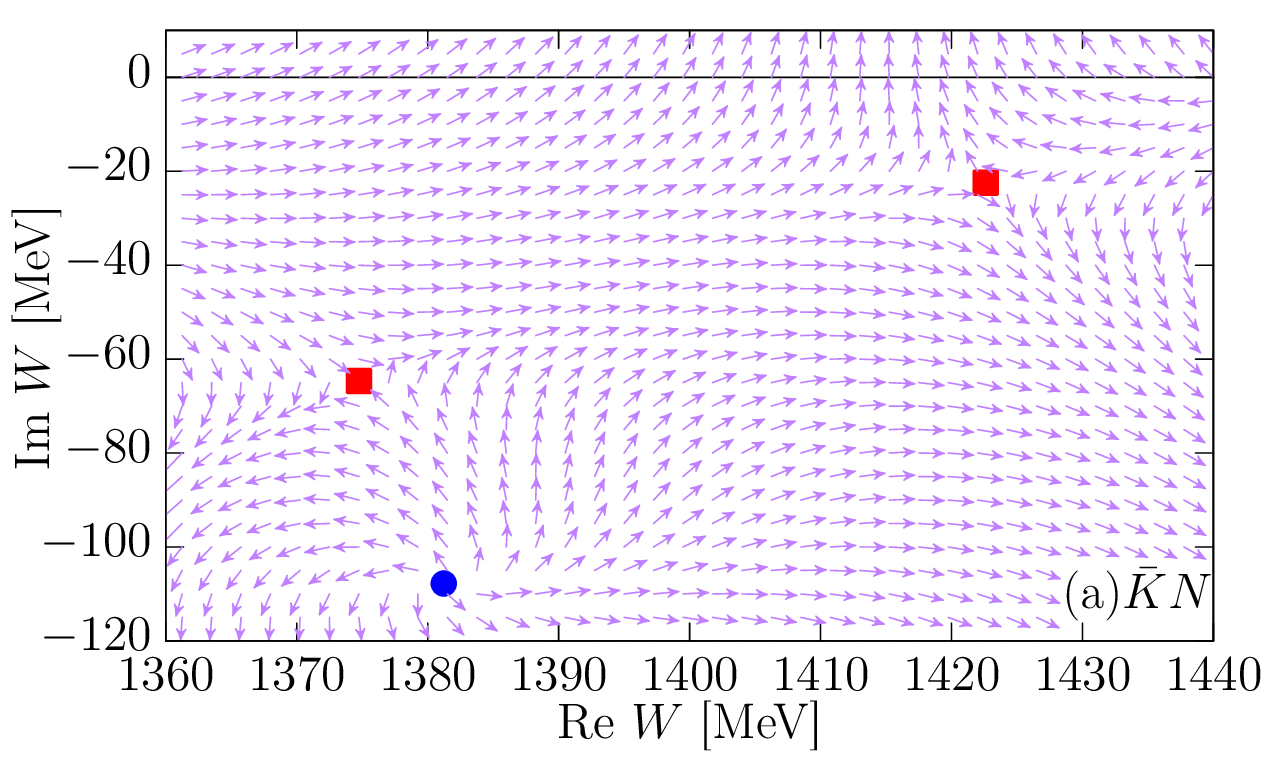}
	\includegraphics[width=75mm]{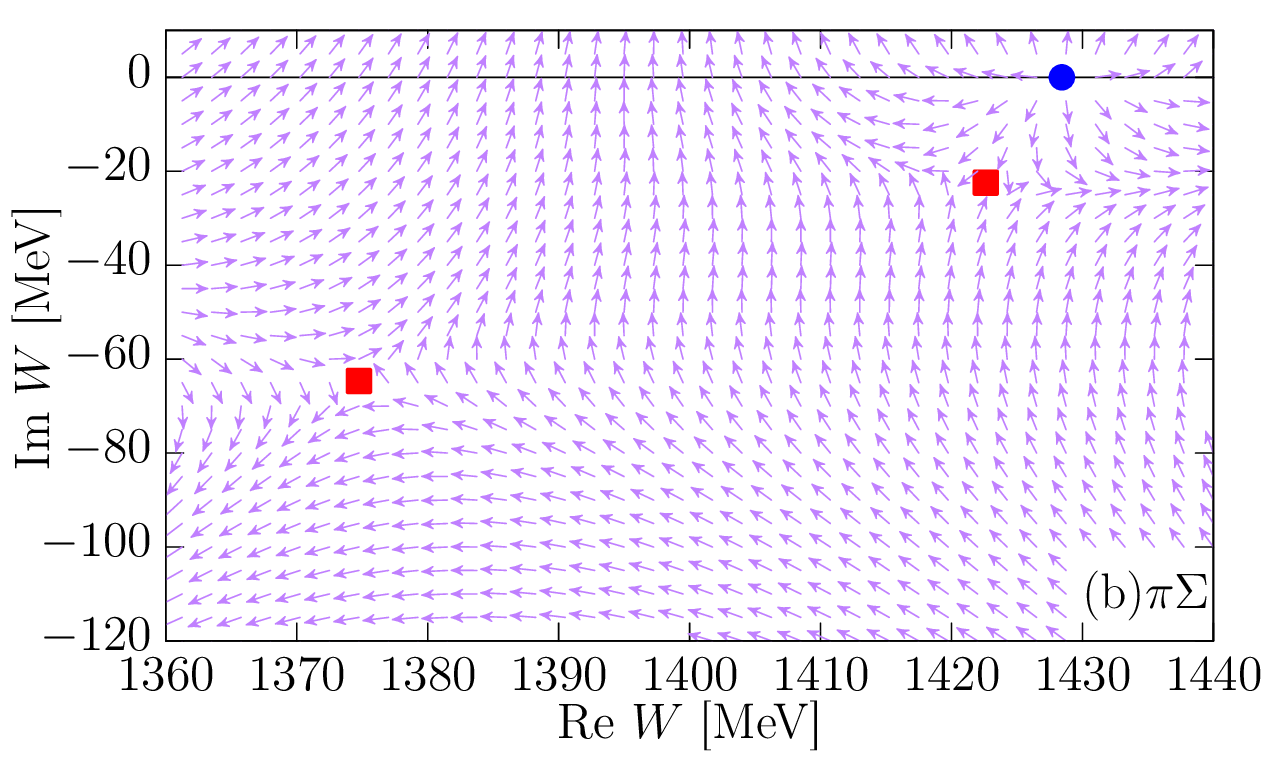}
	\caption{Positions of the poles (squares) and CDD zeros (circles) in the $\bar{K}N$ amplitude (a) and the $\pi\Sigma$ amplitude (b). The angle of vectors from the real axis denotes	the phase of amplitude $\arg\mathcal{F}_{ii}(z)$.}
	\label{fig:vector_map}
\end{figure}

To study the $\bar{K}N$-$\pi\Sigma$ scattering amplitude, 
we use the effective Tomozawa-Weinberg (ETW) model~\cite{Ikeda:2012au}
in which the experimental data around the $\bar{K}N$ threshold energy is well reproduced by the Tomozawa-Weinberg term with the $\bar{K}N$, $\pi\Sigma$  and $\pi\Lambda$ channels.
The Tomozawa-Weinberg interaction is an energy-dependent four-point contact interaction that is the leading order term in chiral perturbation theory.
In the ETW model, two poles of $\Lambda(1405)$ are also dynamically generated in the coupled- channel scattering amplitude.
With the isospin symmetric hadron masses, the positions of the low-mass and high-mass poles are,  respectively, found to be 
\begin{align}
W^{\mathrm{Low}}_\mathrm{pole}=1375-65i\physdim{MeV},\quad  W^{\mathrm{High}}_{\mathrm{pole}}=1423-22i\physdim{MeV}.
\end{align}

\begin{figure}[t]
	\centering
	\includegraphics[width=75mm]{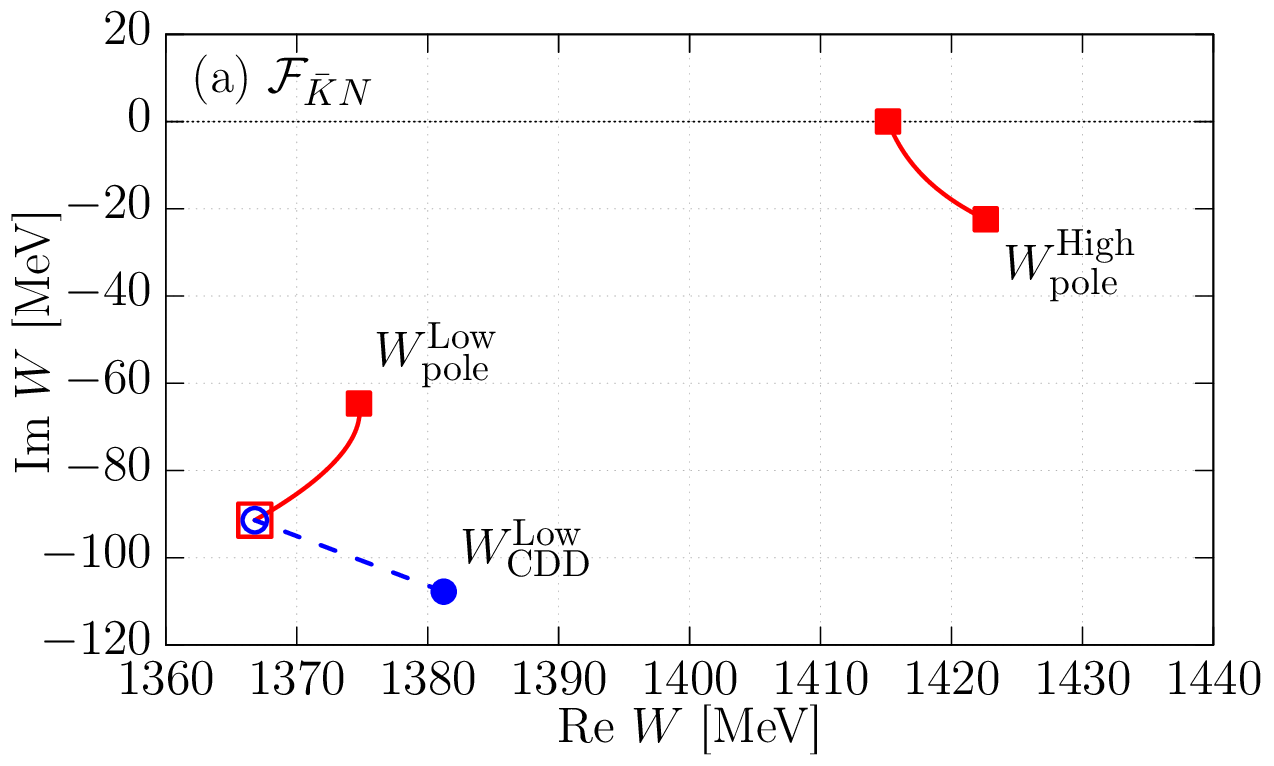}
	\includegraphics[width=75mm]{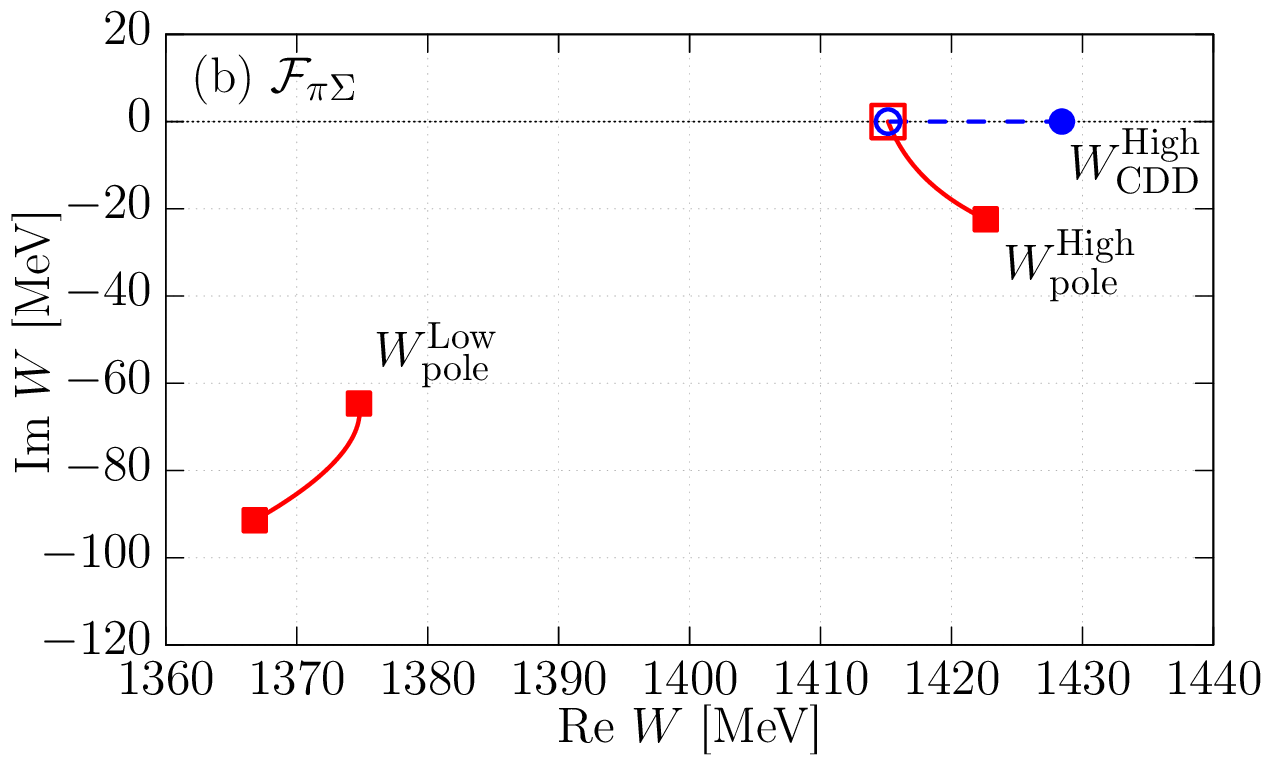}
	\caption{Trajectories of poles (solid line) and CDD zeros (dashed line) in the $\bar{K}N$ amplitude (a) and the $\pi\Sigma$ amplitude (b).}   
	\label{fig:trajectory_Lambda}
\end{figure}

Searching for the $\bar{K}N$  and $\pi\Sigma$ amplitudes,
we find one CDD zero in each component.
Because the CDD zero in the $\bar{K}N$ ($\pi\Sigma$) channel lies near the low-mass (high-mass) pole,
we denote its energy as $W_{\mathrm{CDD}}^{\mathrm{Low}}$ ($W_{\mathrm{CDD}}^{\mathrm{High}}$).
The positions of the zeros are 
\begin{align}
W_{\mathrm{CDD}}^{\mathrm{Low}} = 1381-108i\physdim{MeV}, \quad W_{\mathrm{CDD}}^{\mathrm{High}} = 1428-0i\physdim{MeV}.
\end{align}
The poles and CDD zeros are shown in Fig.~\ref{fig:vector_map} together with the phase of the amplitude.
It is seen that the phase increases $2\pi$ (decreases $2\pi$) along with the contour enclosing the CDD zero (pole).
This rotation of the phase gives the topological invariant $n_C$ in Eq.~\eqref{eq:nC}.
The high-mass pole is accompanied by the CDD zero in the $\pi\Sigma$ channel, but not in the $\bar{K}N$ channel.
This means that the high-mass pole originates in the $\bar{K}N$ channel.
This result is consistent with the dominance of the $\bar{K}N$ composite component in this eigenstate as suggested in various studies~\cite{Jido:2003cb,Hyodo:2007jq,Sekihara:2014kya,Miyahara:2015uya,Kamiya:2016oao,Guo:2015daa,Kamiya:2015aea}.
In the same manner, the origin of the low-mass pole is in the $\pi\Sigma$ channel as indicated in Refs.~\cite{Jido:2003cb,Hyodo:2007jq,Guo:2015daa}

In the ETW model, the ZCL is achieved by suppressing the off-diagonal interaction $V_{\bar{K}N\mathchar`-\pi\Sigma}\rightarrow 0$, keeping the diagonal parts $V_{\bar{K}N\mathchar`-\bar{K}N}$  and $V_{\pi\Sigma\mathchar`-\pi\Sigma}$ unchanged.
The trajectories of the poles and CDD zeros toward the ZCL are shown in Fig.~\ref{fig:trajectory_Lambda}.
The poles appear at the same position in each component so that the trajectories are also identical with each other.
We see that the high-mass pole moves toward $W=1415\physdim{MeV}$ on the real axis. 
In the ZCL, this pole decouples from the $\pi\Sigma$  amplitude, while it remains in the $\bar{K}N$ amplitude as a bound state pole~\cite{Hyodo:2007jq}.
Because the diagonal parts of the interaction are unchanged in the ZCL,
this pole is dynamically generated by the single-channel $\bar{K}N$ interaction.
In contrast to this, the $\pi\Sigma$ pole remains in the $\pi\Sigma$ amplitude as a single-channel resonance at $W=1367-91i\physdim{MeV}$, but it decouples from the $\bar{K}N$ amplitude.
These trajectories of poles are consistent with the results in Ref.~\cite{Cieply:2016jby}.

Now let us consider the trajectories of the CDD zeros.
In the $\bar{K}N$ amplitude, we see that the low-mass zero $W_{\mathrm{CDD}}^{\mathrm{Low}}$ 
encounters with the low-mass pole and decouples from the amplitude in the ZCL.
The high-mass zero in the $\pi\Sigma$ amplitude moves toward the real axis and annihilates the high-mass pole.
This result shows that the physical amplitude is close to the ZCL
because the modification of the eigenenergy (a few tens of $\mathrm{MeV}$) is much smaller than the mass of the eigenstate.
Thus, as expected by the positions of the poles and CDD zeros, the $\pi\Sigma$ ($\bar{K}N$) dynamical  component is not dominant in the structure of the state represented by the high-mass (low-mass) pole.

\section{Summary}
We have proposed a useful method to study the origin of hadron resonances. It is shown that the eigenstate pole should be accompanied by a nearby CDD zero, if the resonance originates in the dynamics of a hidden channel. The existence of the zero is robust, as it is topologically guaranteed by Eq.~\eqref{eq:nC}. Moreover, a CDD zero near the pole causes the distorted line shape of the resonance from the Breit-Wigner form as an observable consequence. Applying this method to $\Lambda(1405)$, we show that the high-mass (low-mass) pole originates in the $\bar{K}N$ ($\pi\Sigma$) dynamics from the position of the CDD zeros in the $\pi\Sigma$ ($\bar{K}N$) amplitude.

We summarize how to apply our method in practice.
One way is to find the distortion of the peak of the invariant mass distribution.
As discussed in Sec. IV, the existence of the CDD zero distorts the line shape of the peak from the Breit-Wigner form.
While there are various mechanisms that distort the line shape,
the observation of the distorted peak may give us a hint to search for a possible CDD zero near the eigenstate pole.
A better way is to extract the two-body scattering amplitude by analyzing the experimental data of final state interactions with sufficiently accurate data as in the $\pi\pi$ scattering.
In this case, we can apply our method directly, by searching for the position of the pole and CDD zero of the fitted scattering amplitude.
Because of its generality, the method developed here will shed light on the origin of many hadron resonances.

\acknowledgments

This work is in part supported 
by JSPS KAKENHI Grants No.　JP17J04333 and No. JP16K17694 and by the Yukawa International Program for Quark-Hadron Sciences (YIPQS).

\end{document}